\documentclass{jpsj3}  
\usepackage{txfonts}

  \title{
Spontaneous Magnetic Field near a Time-Reversal 
Symmetry Broken Surface State of YBCO
}
 
\author{Kazuhiro Kuboki \thanks{kuboki@kobe-u.ac.jp}}
\inst{Department of Physics, Kobe University, Kobe 657-8501, Japan
} 

\abst{
Spatial distributions of spontaneous magnetic fields 
near a surface of cuprate high-$T_C$ superconductor YBCO 
with broken time-reversal symmetry are calculated using 
the Ginzburg-Landau theory derived from the $t-J$ model. 
It is found that the magnetic field exists in a 
narrow region inside the superconductor, and it decays quickly 
outside the surface. 
Experimental approaches possible to detect such 
spontaneous magnetic fields are discussed. 
}

\begin{document}
\maketitle

\newpage

More than two decades ago, the peak splitting of zero bias 
conductance in $ab$-oriented 
YBCO/insulator/Cu tunnel junction was observed,\cite{Coving} 
and it has been considered as a sign of spontaneous violation of 
time-reversal symmetry (${\cal T}$).\cite{SigRev} 
To interpret this experiment, superconducting (SC) states with 
a second SC order parameter (OP) that has symmetry different 
from that in the bulk ($d_{x^2-y^2}$ wave) have been 
proposed.\cite{Matsu1,Matsu2,Fogel}
In these states, spontaneous currents and 
magnetic fields would occur near the surface. 
However, their existence is still controversial.\cite{Carmi,msr}

The present author studied (110) surface states of 
YBCO that has two CuO$_2$ planes in a unit cell, using 
a bilayer $t-J$ model and the Bogoliubov de Gennes (BdG) 
method.\cite{KK3,KK4} 
Near the (110) surface, where the $d_{x^2-y^2}$-wave SCOP 
($\Delta_d$) is strongly suppressed, flux phase appears locally 
leading to a ${\cal T}$-breaking surface state. 
Flux phase is a mean-field solution to the $t-J$ model in which 
staggered currents flow and the flux penetrates a plaquette in 
a square lattice.\cite{Affleck} 
Although this state is only a metastable solution, its free energy 
is close to that of $d_{x^2-y^2}$-wave SC state,  
because the former also has $d_{x^2-y^2}$ symmetry.\cite{Bejas,KK1} 
(On the contrary, on the basis of the $t-J$ model, 
$s$-wave SC state is not favored when the $d$-wave state is realized.) 
Then, the flux phase can occur once the $d$-wave 
SC order is suppressed.
In the surface state with local flux phase order, the current  
is oscillating as a function of the distance from 
the surface. 
Moreover, in the bilayer model the fluxes in two layers are 
opposite.\cite{Zhao,KK3} 
Thus the spontaneous magnetic field near the surface 
is expected to be small. 

In this short note, we study the spatial distributions of spontaneous 
magnetic fields near the (110) surface of YBCO, 
using the Ginzburg-Landau (GL) free energy derived from 
the $t-J$ model.\cite{KKGLFL}
The purpose is to show that it will be difficult to observe 
the spontaneous field outside the sample even when ${\cal T}$ 
is broken, while it may be detected inside the superconductor. 
The  GL theory in this work is derived from the microscopic model 
and the parameters are chosen to represent the electronic 
structure of YBCO. This is the difference from other theories 
that have been proposed to explain the absence of 
spontaneous magnetic fields.\cite{Lee2,Fogel2}

First we consider the single layer case. 
GL free energy $F_{GL}$ derived from the $t-J$ model consists of 
SCOPs ($\Delta_d$ and $\Delta_s$), flux phase OP ($\Pi$), 
and the vector potential. 
Coefficients of all terms in $F_{GL}$ can be calculated 
as functions of the first- ($t$), second- ($t'$), and third- ($t''$) 
neighbor transfer integrals, the superexchange interaction ($J$), 
the doping rare ($\delta$), and the temperature ($T$). 
We take $x$ ($y$) axis perpendicular (parallel) to the 
(110) surface. 
The region $x > 0$ ($x < 0$) is a superconductor (vacuum), 
and we assume that the system is uniform along the $y$ direction. 
Numerical solutions are obtained by applying a quasi-Newton method 
to $F_{GL}$ under the constraint that $\Delta_d$ vanishes 
at $x=0$. 

In Fig. 1, the spatial variations of the OPs are shown. 
Here we take $t/J=2.5 \ (J=0.1$eV), $t'/t=-0.3$, $t''/t=0.15$, 
$\delta=0.13$, and $T=0.45T_C$ ($T_C=0.131J$ being 
the SC transition temperature) and 
$\xi_d$ is the coherence length of $\Delta_d$. 
(For the parameters used here and the lattice constant of the 
 square lattice $a=3.8$ \AA, $\xi_d \sim $9.4 \AA .) 
It is seen that the flux phase OP becomes finite near the surface 
where the $d$-wave SCOP is suppressed. 
In this region spontaneous staggered currents $J_y$ 
and magnetic fields appear.
%
\begin{figure}[htb]
\begin{center}
\includegraphics[width=7.0cm,clip]{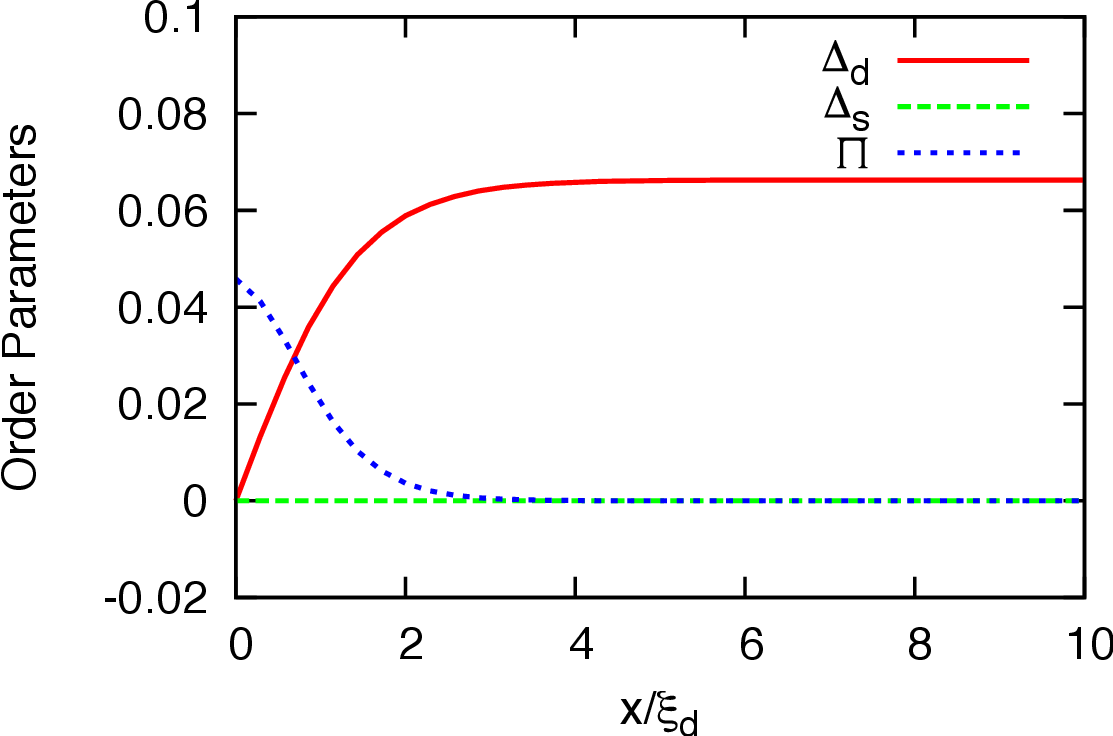}
\caption{(Color online) Spatial variations of the $d$-wave ($\Delta_d$)  
and $s$-wave ($\Delta_s$) SCOPs,  and the OP for the flux phase ($\Pi$) 
near the (110) surface. Note that all OPs are non dimensional.} 
\end{center}
\end{figure}

Next we treat a bilayer system by stacking the single-layer systems 
in such a way that the fluxes (and the currents) point 
oppositely in neighboring layers. 
The $c$ axis lattice constant 
and the distance between a bilayer are taken to be $c=11.7$ \AA \
and $c_1=3.4$ \AA, respectively, to represent the structure of YBCO. 
We take the origin of the $z$ axis at the center of a bilayer, 
and consider infinite stacking of CuO$_2$ planes. 
(Distance between neighbor currents in a plane is 
$a/\sqrt{2}$.\cite{KK2,KK4})
We calculate magnetic fields ${\bf B}$ by using the 
Biot-Savart law 
with the spontaneous currents obtained in the GL solutions 
to the single layer case.
In this approach the screening effects in superconductors 
are neglected, so that the results should be taken as an upper limit 
of the absolute value of $B$. 

Now we show the spatial variations of the magnetic fields in Fig.2. 
It is seen that both in-plane ($B_x$) and vertical ($B_z$) components 
occur near the surface, $x \lesssim \xi_d$, 
and they are oscillating as functions of $x$.
(For $z=0$, $B_z$ vanishes due to symmetry.)
%
\begin{figure}[htb]
\begin{center}
\includegraphics[width=7.0cm,clip]{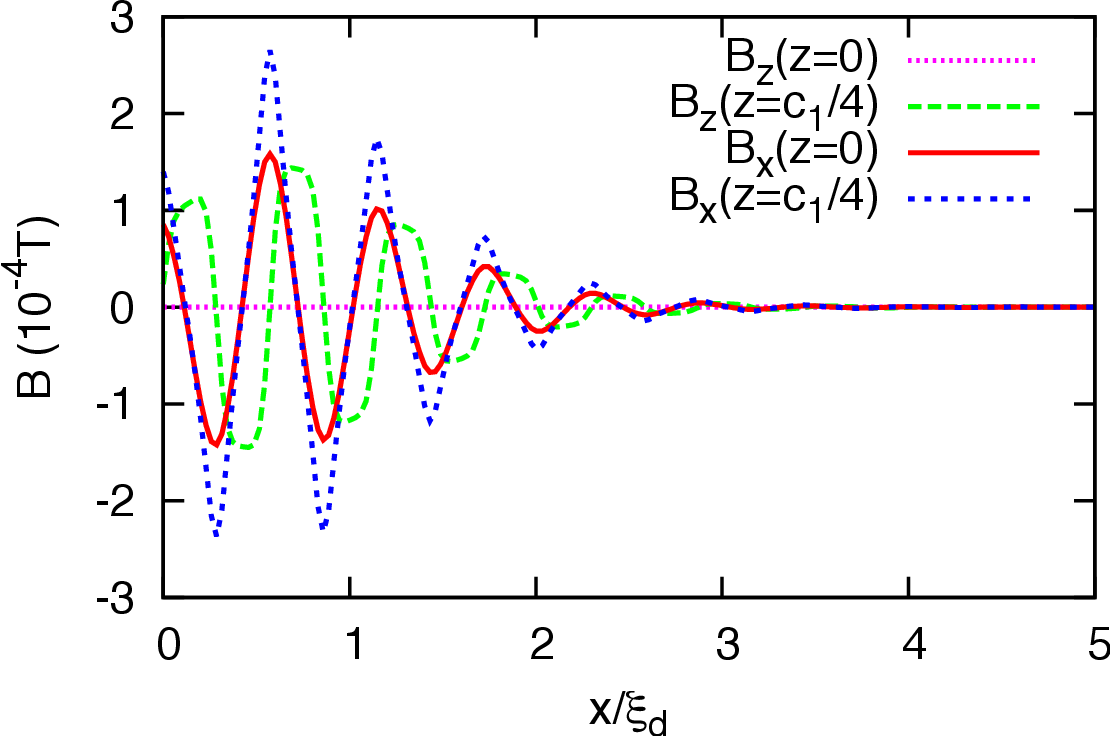}
\caption{(Color online) Spatial variations of $B_z$ and $B_x$  
as functions of the distance from the surface, $x$, for 
$z=0$ and $z=c_1/4$.}
\end{center}
\end{figure}

In fig.3, $B_z$ and $B_x$ are presented as functions of $z$ for 
two choices of $x \ (<0)$. (The points $z=0$ and $z=\pm c$ are 
the center of a bilayer in neighboring unit cell.) 
Outside the surface, both $|B_z|$ and $|B_x|$ are quite small. 
For $x=-0.2c$, $B_x$ and $B_z$ are already $\sim$ 1\% 
compared to those inside the superconductor, 
and they almost vanish for $x=-c$. 
%
\begin{figure}[htb]
\begin{center}
\includegraphics[width=7.0cm,clip]{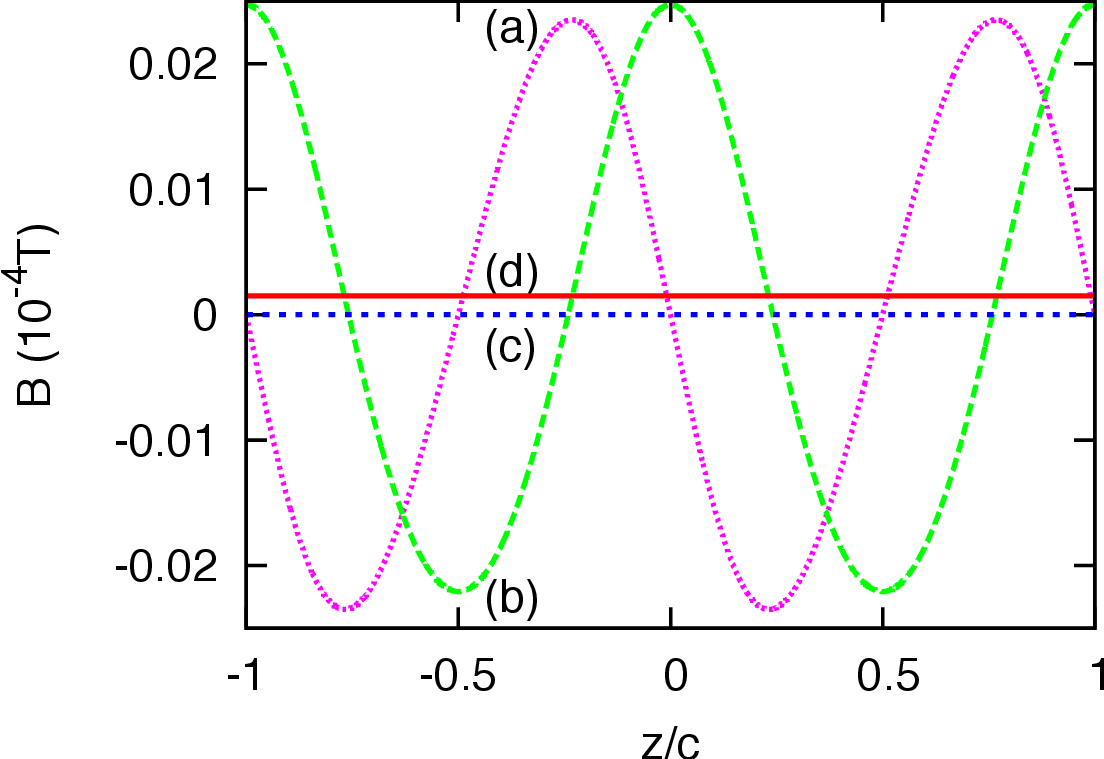}
\caption{(Color online) z dependence of $B_z$ and $B_x$ outside 
the superconductor. 
(a) $B_z(x=-0.2c,z)$, (b) $B_x(x=-0.2c,z)$, 
(c) $B_z(x=-c,z)\times10^2$, and (d) $B_x(x=-c,z)\times10^2$.}
\end{center}
\end{figure}
%
For the sake of comparison, we show the results for the system 
where the currents in all layers are parallel. (Fig.4)
%
\begin{figure}[htb]
\begin{center}
\includegraphics[width=7.0cm,clip]{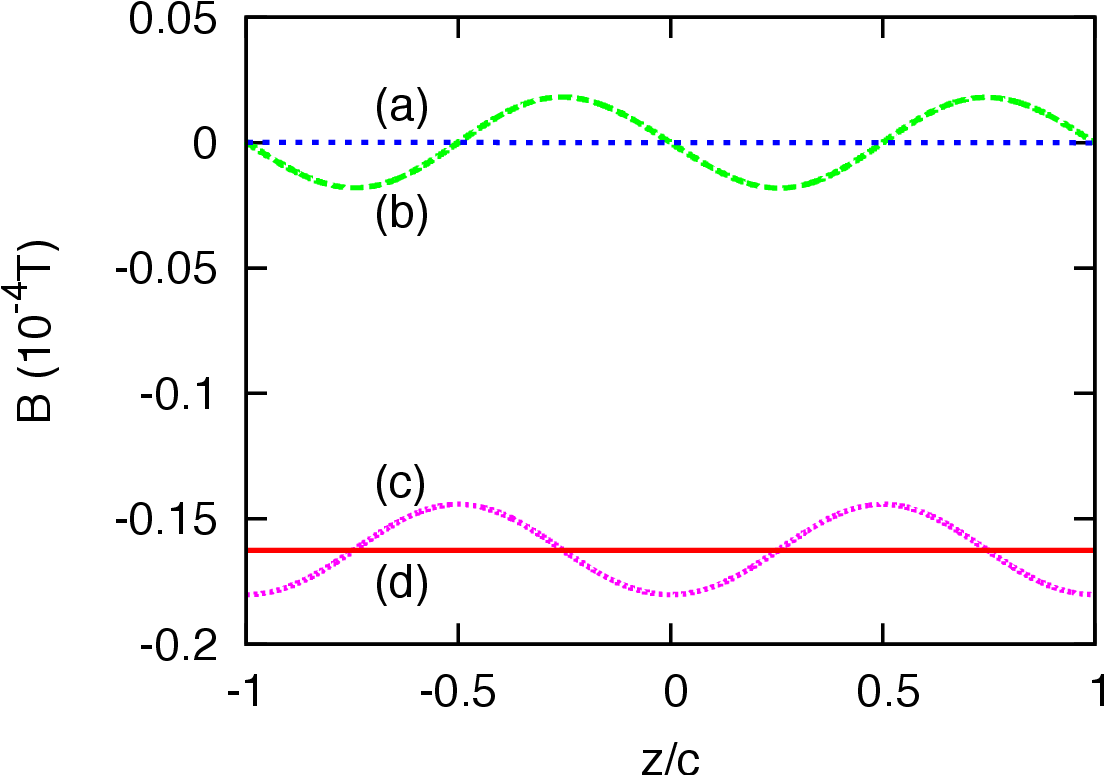}
\caption{(Color online) z dependence of $B_z$ and $B_x$ outside 
the superconductor, when the currents in neighboring layers 
are parallel.
(a) $B_x(x=-100c,z)$, (b) $B_x(x=-0.2c,z)$, 
(c) $B_z(x=-0.2c,z)$, and (d) $B_z(x=-100c,z)$.}
\end{center}
\end{figure}
%
In this case $B_z$ stays constant when $x \gg c$, while $B_x$ 
becomes zero. 
This is because the situation is equivalent to 
uniform currents flowing in an infinite plane ((110) plane), 
if one looks from a point far away from the surface. 
In the case of antiparallel currents, however, 
cancellation among contributions from staggered currents 
forces ${\bf B}$ to vanish. 
Then the typical length scale for the decay of ${\bf B}$ is of 
the order of  $c_1$, i.e., the distance between bilayer, or, 
antiparallel currents. 

The difference between the cases of parallel and antiparallel 
currents is crucial to explain the apparent absence of 
spontaneous fields at the (110) surface of YBCO, where 
the peak spilling of the zero bias conductance 
was observed and ${\cal T}$ breaking has been expected. 
Since the spontaneous magnetic field exists essentially 
inside the sample, it would be difficult to detect it 
using, e.g., SQUID microscope. 
Experimental approaches possible to measure it may be 
$\mu$SR or polarized neutron scattering. 
 
In the scenario to explain ${\cal T}$ violation using 
the second SCOP, e.g., $(d\pm is)$-wave state, spontaneous 
currents on different layers would be parallel,  
because Josephson coupling between layers should  
favor the phase difference to be zero.
Thus ${\bf B}$ may be observable outside the system 
in this case. 

In this work, we obtained the absolute value of $B$ 
(inside the superconductor) of the order of 1 G.  
However, it could be larger if $T$ is lowered. 
Since the GL theory cannot handle the temperature 
region $T \ll T_C$, other approach, e.g., 
the BdG method has to be employed to examine this issue. 
We will study this problem separately.

\begin{acknowledgment}
 The author thanks M. Hayashi for useful discussions. 
\end{acknowledgment}


\end{document}